\documentclass[11pt]{article}

\RequirePackage{latexsym}
\RequirePackage{graphicx}
\RequirePackage{amssymb}
\RequirePackage{natbib}
\RequirePackage{setspace}

\usepackage{color}
\usepackage{array}
\usepackage{hyphenat}
\usepackage{fancyvrb}
\usepackage{fixltx2e}

\newcolumntype{q}[1]{>{\setlength{\parindent}{1em}}p{#1}}

\usepackage{fancyhdr}
\begin{document}
\definecolor{gold}{rgb}{0.85,0.66,0}
\definecolor{grey}{rgb}{0.5,0.5,0.5}
\definecolor{blue}{rgb}{0,0,1}
\definecolor{red}{rgb}{1,0,0}
\definecolor{green}{rgb}{0,0.7,0}
\definecolor{green}{rgb}{0,0,0}

     \begin{figure}
     \begin{center}
     \includegraphics[scale=1]{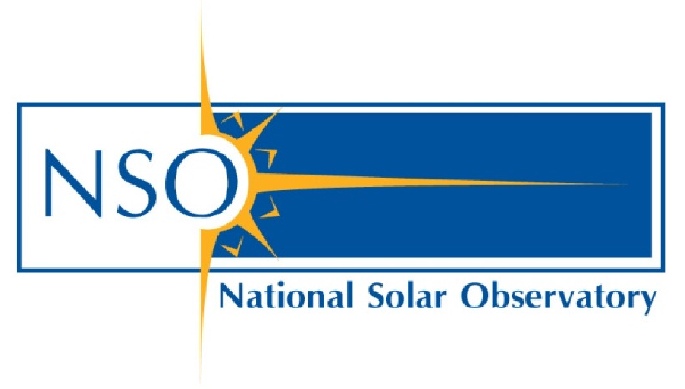}
     \end{center}
     \end{figure}
\title{{\bf HMI Synoptic Maps Produced by NSO/NISP}
\\$\,$}
\author{Anna L.~H.~Hughes, Luca Bertello, Andrew R.~Marble, Niles A.~Oien, \\Gordon Petrie, and Alexei A.~Pevtsov
\\$\,$
\\ National Solar Observatory
\\$\,$  \\$\,$  \\$\,$}
\maketitle
\thispagestyle{empty}

\noindent\rule{\linewidth}{0.2mm}
\rule{\linewidth}{0.5mm}

\begin{center}
Technical Report No. {\bf NSO/NISP-2016-002}
\\$\,$
\end{center}

\begin{abstract}

\textcolor{green}{Recently, the National Solar Observatory (NSO) Solar-atmosphere Pipeline Working Group (PWG) has undertaken the production of synoptic maps from Helioseismic and Magnetic Imager (HMI) magnetograms.  A set of maps has been processed spanning the data available for 2010-2015 using twice daily images (taken at UT midnight and noon) and running them through the same algorithms used to produce SOLIS/VSM 6302l mean-magnetic and spatial-variance maps.  The contents of this document provide an overview of what these maps look like, and the processing steps used to generate them from the original HMI input data.}

\end{abstract}
$\,$

\pagebreak
\tableofcontents
$\,$\\
$\,$

\section[\textcolor{green}{Basic Product: Integral Carrington Maps}]{\textcolor{green}{Basic Product: Integral Carrington Maps}}
\label{Product}

The goal of this project has been to create a series of integral magnetic synoptic maps using HMI data \citep{citeHMI_instrument} and run using the same algorithms as those that produce the spatial-variance synoptic maps outlined in \cite{citeBertello2014} and derived from NSO SOLIS (Synoptic Optical Long-term Investigations of the Sun) VSM (Vector Spectromagnetograph) 630.2 nm data.

The HMI synoptic maps that we have produced are all integral synoptic maps (spanning 0-360$^\circ$ of a single Carrington rotation), in Carrington-longitude--by--sine(latitude) binning, and in NSO-low-resolution format (360x180 map bins, where contributing observations are weighted by cosine$^4$($\Delta$longitude) relative to the observed longitude of central meridian).

Unlike the SOLIS/VSM spatial-variance maps produced to date, the HMI synoptic maps have been run using both longitudinal- and vector-observed magnetograms, where the final data-products for each are detailed in \S\ref{Product_LOS} and \S\ref{Product_Vector}, below.  At a minimum, each FITS-file map set includes a frame for:

- the mean photospheric radial magnetic flux, 

- the spatial variance of the mean photospheric radial magnetic flux, 

- the sum-of-weights from all observations contributing to a given map set.

	\subsection[\textcolor{green}{Pseudo-radial Maps from LOS Data}]{\textcolor{green}{Pseudo-radial Maps from LOS Data}}
	\label{Product_LOS}

For the HMI-longitudinal synoptic maps, we have used data from the HMI m\_720s series (longitudinal magnetograms covering a 12-minute integration window, \cite{citeHMI_basic}).  In order to produce maps of radial magnetic flux, we have projected the line-of-sight flux values into pseudo-radial values using the assumption of a perfectly radial magnetic field at the photosphere:
\begin{equation}
   B_{r,\mathrm{pseudo}}
=
   B_{\mathrm{LOS}} / \cos\left(\rho\right)
   \,,
\label{EQ_radproj}
\end{equation}
where $B_\mathrm{LOS}$ is the observed LOS flux, and $\rho$ is the center-to-limb (or heliocentric) angle between the line-of-sight vector and the local vertical.

Additionally, with the reasonable levels of quiet-sun sensitivity provided by longitudinal observations, basic methods for filling in unobservable or poorly observed polar fields become viable.  Therefore, for the HMI-LOS derived maps, we have provided a pole-filled version of the mean--pseudo-radial--flux map as an additional frame.  Some methods of pole-filling interpolate spatially and temporally across the pole from well-observed dates/latitudes \citep{citeSun2011}.  In our case, the polar fields are filled in using a cubic-polynomial surface fit to the currently observed fields at neighboring latitudes. The fit is performed on a polar-projection of the map using low standard-deviation-to-fit measurements only, and the high-latitude fit is then integrated into the observed synoptic map, weighting toward the pole.

\begin{figure}
     \begin{center}
     \includegraphics[scale=0.19]{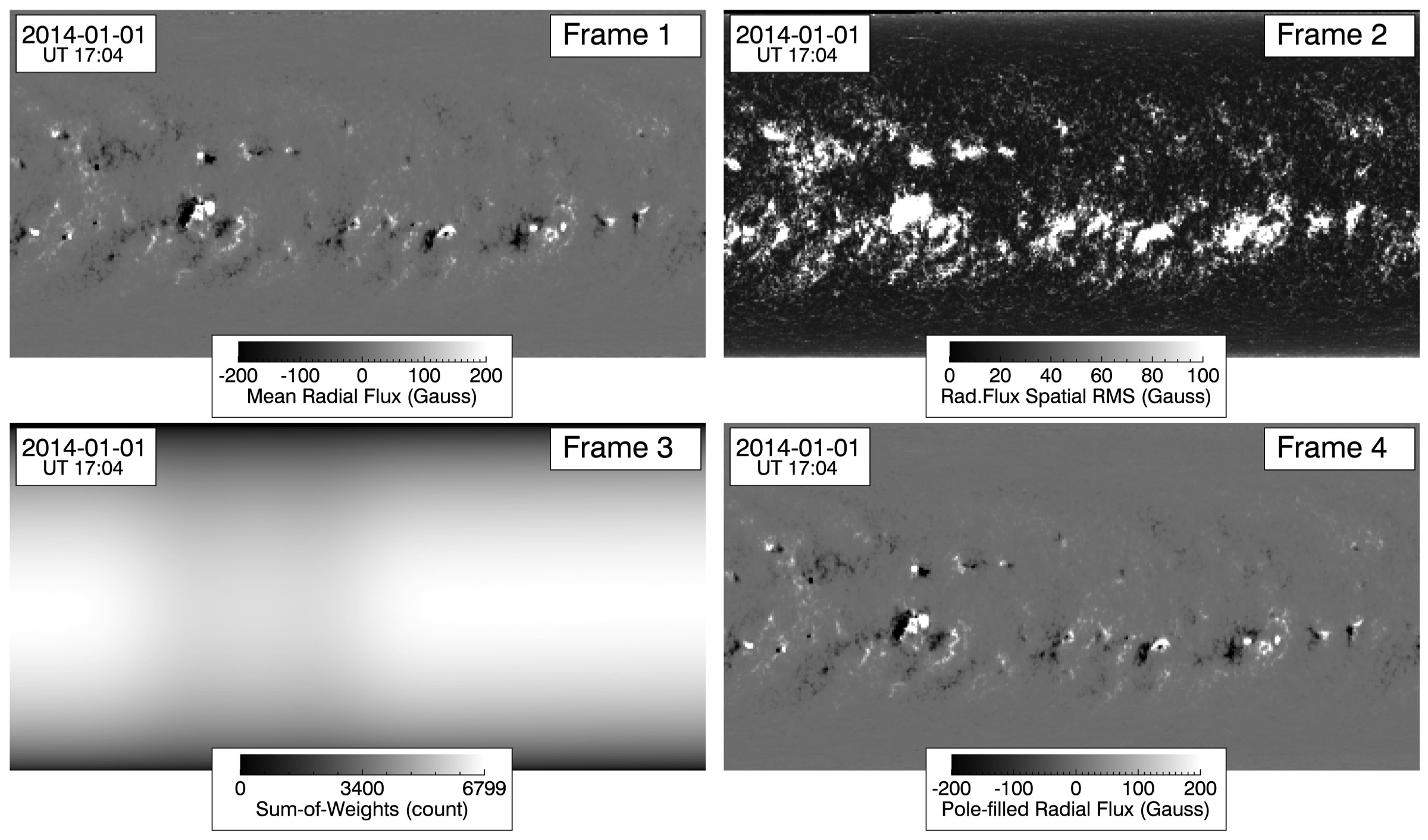}
     \caption[Example LOS Synoptic Map]{Set of example image frames for an HMI-LOS-derived synoptic map: Carrington rotation 2145 centered, on April 1$^\mathrm{st}$, 2014.}
     \label{FIG_Product_LOSsynoptic}
     \end{center}
     \end{figure}

A set of example maps derived from HMI-longitudinal magnetograms is shown in Figure~\ref{FIG_Product_LOSsynoptic},
while the file-name structure that we have used and the specifics of the FITS-file frame contents are outlined below.\\

\noindent{\bf Filename Structure:}\\
`xbx73{\bf YYMMDD}t{\bf HHMM}c{\bf CCCC}\_000\_int-err\_dim-180\_source-SDO-HMI.fits.gz'
\begin{list}{$\circ$}{}
   \item `xbx73':  This is the product code that denotes HMI synoptic maps derived from photospheric longitudinal magnetograms.
   \item `{\bf YYMMDD}t{\bf HHMM}':  This is the time-stamp assigned to the map.  For Integral synoptic maps, NSO uses the date and time corresponding to the midpoint of a given Carrington rotation.
   \item `c{\bf CCCC}\_000':  This denotes the Carrington rotation mapped as well as the Carrington longitude at the left map edge.  As Integral synoptic maps always run from 0 to 360 degrees, the filenames for these maps will always have `\_000' for the longitude.
\end{list}

\pagebreak
\noindent {\bf FITS-frame Contents:}\\
$\,$\\
\noindent
\begin{tabular}{cp{0.6in}q{4.9in}}
     {\bf frame} &
        {\bf units} &
        \noindent {\bf title / description}
        \\
     \hline
     &&\\
     1 & Gauss &
        \noindent {\bf Weighted-mean Radial Flux Density:}\endgraf
        $\,\,\,\,\,\,\,\,\,$The mean value of the radially-projected HMI-LOS magnetograms for each longitude-sine(latitude) map bin.  Each input observation is spatially weighted to emphasize contributions observed near the central meridian. 
        \\
        &&\\
     2 & Gauss &
        \noindent {\bf Spatial RMS Estimate:} \endgraf
        $\,\,\,\,\,\,\,\,\,$The weighted statistical variance of all radially-projected HMI-LOS-magnetogram-values contributing to a given longitude-sine(latitude) bin (corresponding to the mean-flux values in Frame 1). 
        \\
        &&\\
     3 & counts &
        \noindent {\bf Sum-of-Weights:} \endgraf
        $\,\,\,\,\,\,\,\,\,$The sum of weights into each map bin.  This includes both the $\Delta$longitude--versus--central-meridian weighting applied across each input observation, as well as the count of sky-image pixels contributing to each observed longitude-sine(latitude) bin.
        \\
        &&\\
     4 & Gauss &
        \noindent {\bf Pole-filled Mean Radial Flux Density:} \endgraf
        $\,\,\,\,\,\,\,\,\,$The pole-filled version of Frame 1. 
        \\
        &&\\
\end{tabular}


	\subsection[\textcolor{green}{3-Component Maps from Vector Data}]{\textcolor{green}{3-Component Maps from Vector Data}}
	\label{Product_Vector}

For the HMI-vector synoptic maps, we have used data from the HMI b\_720s series (fully disambiguated vector magnetograms, \cite{citeHMI_vectorpipe}), choosing to apply the results of the Radial-accute disambiguation for the regions of quiet sun \citep{citeMetcalf2006,citeLeka2009}.  As this is vector data, the mean-radial-flux map in these files is for true-radial flux.
Additionally, we have mapped the values for the mean poloidal and toroidal fluxes, and computed the spatial-variance of these quantities.  For these additional-component maps, we have used the same cosine$^4$($\Delta$longitude) weighting that broadly emphasizes fluxes observed near central meridian.

\begin{figure}
     \begin{center}
     \includegraphics[scale=0.19]{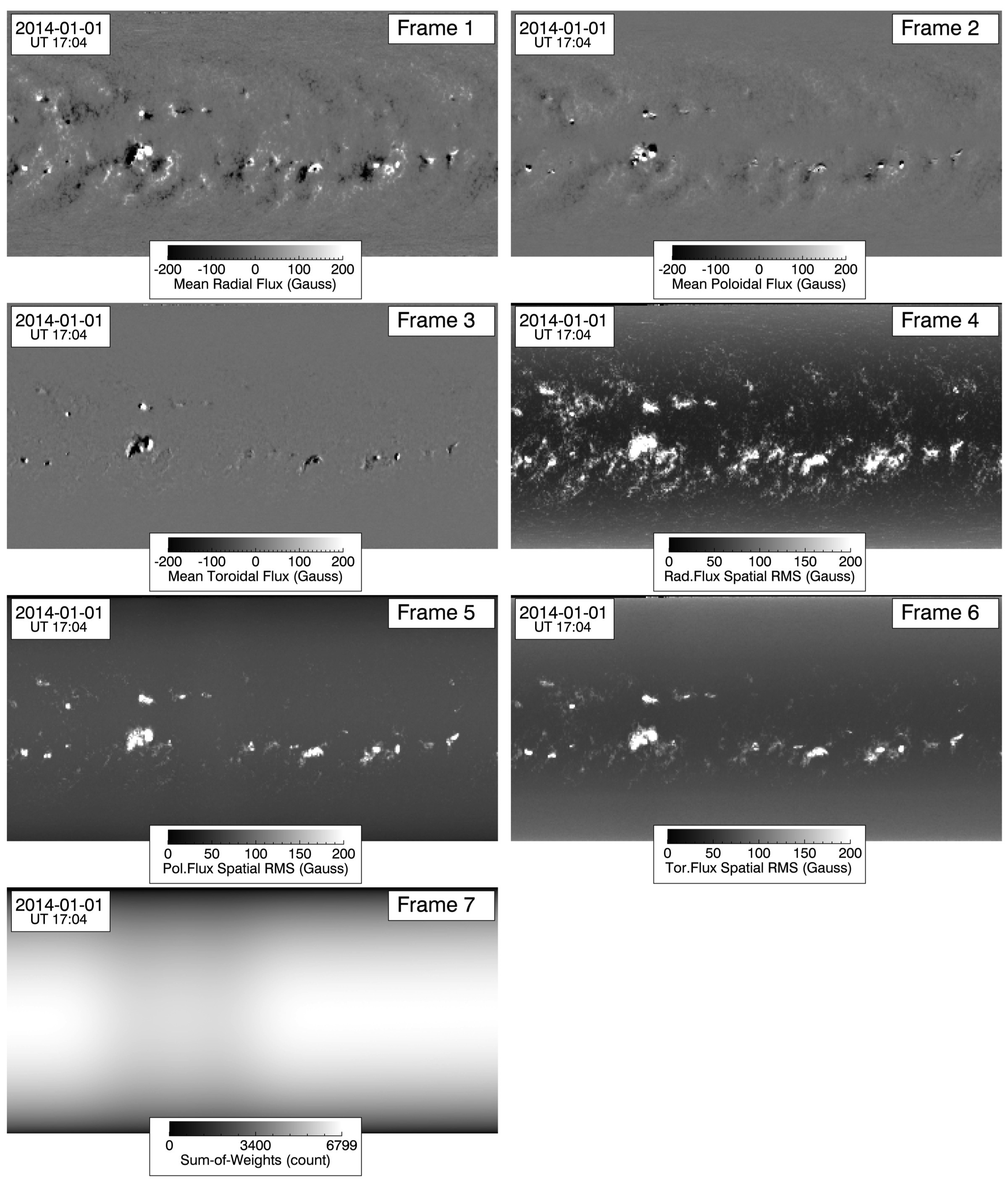}
     \caption[Example Vector Synoptic Map]{Set of example image frames for an HMI-Vector-derived synoptic map: Carrington rotation 2145, centered on April 1$^\mathrm{st}$, 2014.}
     \label{FIG_Product_Vsynoptic}
     \end{center}
     \end{figure}

A set of example maps derived from the HMI-vector magnetograms is shown in Figure~\ref{FIG_Product_Vsynoptic},
while the file-name structure that we have used and the specifics of the FITS-file frame contents are outlined below.\\

\noindent{\bf Filename Structure:}\\
`xbx93{\bf YYMMDD}t{\bf HHMM}c{\bf CCC}\_000\_int-err\_dim-180\_source-SDO-HMI.fits.gz' 
\begin{list}{$\circ$}{}
   \item `xbx93':  This is the product code that denotes HMI synoptic maps derived from photospheric vector magnetograms.
   \item `{\bf YYMMDD}t{\bf HHMM}':  As in \S\ref{Product_LOS}, this is the time-stamp assigned to the map. 
   \item `c{\bf CCCC}\_000':  As in \S\ref{Product_LOS}, this denotes the Carrington rotation mapped as well as the Carrington longitude at the left map edge.
\end{list}

\noindent {\bf FITS-frame Contents:}\\
$\,$\\
\noindent
\begin{tabular}{cp{0.6in}q{4.9in}}
     {\bf frame} &
        {\bf units} &
        \noindent {\bf title / description}
        \\
     \hline
        &&\\
     1 & Gauss &
        \noindent {\bf Weighted-mean Radial Flux Density:} \endgraf
        $\,\,\,\,\,\,\,\,\,$The mean value of the radial (outward) flux (measured from HMI vector magnetograms) for each longitude-sine(latitude) bin.  Each input observation is spatially weighted to emphasize contributions observed near the central meridian.
        \\
        &&\\
     2 & Gauss &
        \noindent {\bf Weighted-mean Poloidal Flux Density:} \endgraf
        $\,\,\,\,\,\,\,\,\,$The mean value of the poloidal (southward) flux (measured from HMI vector magnetograms) for each longitude-sine(latitude) bin.  Each input observation is spatially weighted to emphasize contributions observed near the central meridian.
        \\
        &&\\
     3 & Gauss &
        \noindent {\bf Weighted-mean Toroidal Flux Density:} \endgraf
        $\,\,\,\,\,\,\,\,\,$The mean value of the toroidal (+longitude-ward) flux (measured from HMI vector magnetograms) for each longitude-sine(latitude) bin.  Each input observation is spatially weighted to emphasize contributions observed near the central meridian.
        \\
        &&\\
     4 & Gauss &
        \noindent {\bf Radial-flux Spatial RMS Estimate:} \endgraf
        $\,\,\,\,\,\,\,\,\,$The weighted statistical variance of all HMI-vector radial-flux magnetogram-values contributing to a given longitude-sine(latitude) bin (corresponding to the mean-flux values in Frame 1).
        \\
        &&\\
     5 & Gauss &
        \noindent {\bf Poloidal-flux Spatial RMS Estimate:} \endgraf
        $\,\,\,\,\,\,\,\,\,$The weighted statistical variance of all HMI-vector poloidal-flux magnetogram-values contributing to a given longitude-sine(latitude) bin (corresponding to the mean-flux values in Frame 2).
        \\
        &&\\
     6 & Gauss &
        \noindent {\bf Toroidal-flux Spatial RMS Estimate:} \endgraf
        $\,\,\,\,\,\,\,\,\,$The weighted statistical variance of all HMI-vector toroidal-flux magnetogram-values contributing to a given longitude-sine(latitude) bin (corresponding to the mean-flux values in Frame 3).
        \\
        &&\\
     7 & counts &
        \noindent {\bf Sum-of-Weights:} \endgraf
        $\,\,\,\,\,\,\,\,\,$The sum of weights into each map bin.  This includes both the $\Delta$longitude--versus--central-meridian weighting applied across each input observation, as well as the count of sky-image pixels contributing to each observed longitude-sine(latitude) bin.
        \\
        &&\\
\end{tabular}


\section[\textcolor{green}{Processing Stages and Code Layout}]{\textcolor{green}{Processing Stages and Code Layout}}
\label{Processing}

The following sub-sections provide a basic map of the various processing stages required to ingest HMI sky-image magnetograms and output SOLIS--spatial-variance--style synoptic maps.  These stages include:
\begin{enumerate}
   \item Ingest and prep of HMI sky images (\S\ref{Processing_SkyImages}).
   \item Remapping of sky images into Carrington-longitude--sine(latitude) heliographic maps (\S\ref{Processing_Remaps}).
   \item Combining heliographic remaps into synoptic maps (\S\ref{Processing_Synoptic}).
\end{enumerate}

	\subsection[\textcolor{green}{Ingest of Sky Images}]{\textcolor{green}{Ingest of Sky Images}}
	\label{Processing_SkyImages}

In order to prepare the HMI magnetograms for heliographic and synoptic mapping, a few things need to happen, including: download the magnetograms from the Joint Science Operations Center (JSOC) site, update the image orientation and a few FITS-header keywords to comply with SOLIS-pipeline expectations, and --- in the case of the vector magnetograms --- calculate the heliographic magnetic-vector components from the HMI input frames.  The layout of the code calls looks like this:
\begin{enumerate}
\item Call {\bf backfillMagnetograms.sh} {\em N1 N2}:
     \begin{list}{-}{}
     \item For each day {\em N1} to {\em N2} days ago, requests a download of the UT 00:00 and UT 12:00 magnetograms from JSOC and places the results in an NSO-accessible data-keep directory.
     \end{list}
\item Call {\bf hmi\_serrmaps\_intake2fits\_BatchRun.sh} [-v] {\em START STOP OUTDIR}:
     \begin{list}{-}{}
     \item For each day from {\em START} to {\em STOP}:
          \begin{list}{$\bullet$}{}
          \item Searches for available downloaded JSOC files, {\em INFILE}s.
          \item Calls {\bf hmi\_serrmaps\_intake2fits} [-v] {\em INFILE OUTDIR}
               \begin{list}{{\bf $\,\,\, \rightarrow$}}{}
               \item Opens the rice-compressed {\em INFILE} file.
               \item Rotates the FITS image by 180$^{\circ}$ to place Solar-north at the top.
               \item Re-writes the FITS header using SOLIS-style sectioning.
               \item Adds (primarily duplicate) keywords to the FITS header to account for the updated image geometry and allow for data read-in by SOLIS downstream processing.
               \item Outputs the results to a gzipped file placed in a data keep within {\em OUTDIR} and using the NSO-style file-naming conventions.
               \end{list}
          \end{list}
     \end{list}
\item {\bf IF(vector):}  Call {\bf hmi\_serrmaps\_intakevbundle\_BatchRun.sh} {\em START STOP}:
     \begin{list}{-}{}
     \item For each day from {\em START} to {\em STOP}:
          \begin{list}{$\bullet$}{}
          \item Searches for available ingested gzipped FITS files with flavor tag `\_type-b-720s-field', {\em FIELDFILE}s.
          \item  In IDL, calls {\bf hmi\_serrmaps\_intakevbundle}, {\em FIELDFILE}:
               \begin{list}{{\bf $\,\,\, \rightarrow$}}{}
               \item Using input {\em FIELDFILE} filename to extrapolate, reads in the full file set necessary (`-field', `-inclination', `-azimuth', `-disambig') to compute magnetic-flux vector components.
               \item Applies the disambiguation results (Radial-accute in the quiet sun) to the azimuth image by adding 180$^\circ$ to all pixels where disambig is true.
               \item Calls {\bf chcoord3.pro} to define the heliographic coordinates of each image pixel.
               \item Computes the line-of-site and transverse magnetic-vector components, then rotates them into the local-surface heliographic plane(s).
               \item Outputs the three frames of heliographic vector components into a FITS file with the flavor tag `\_type-b-720s-helio'.
               \end{list}
          \end{list}
     \end{list}
\end{enumerate}

\begin{figure}
     \begin{center}
     \includegraphics[scale=0.19]{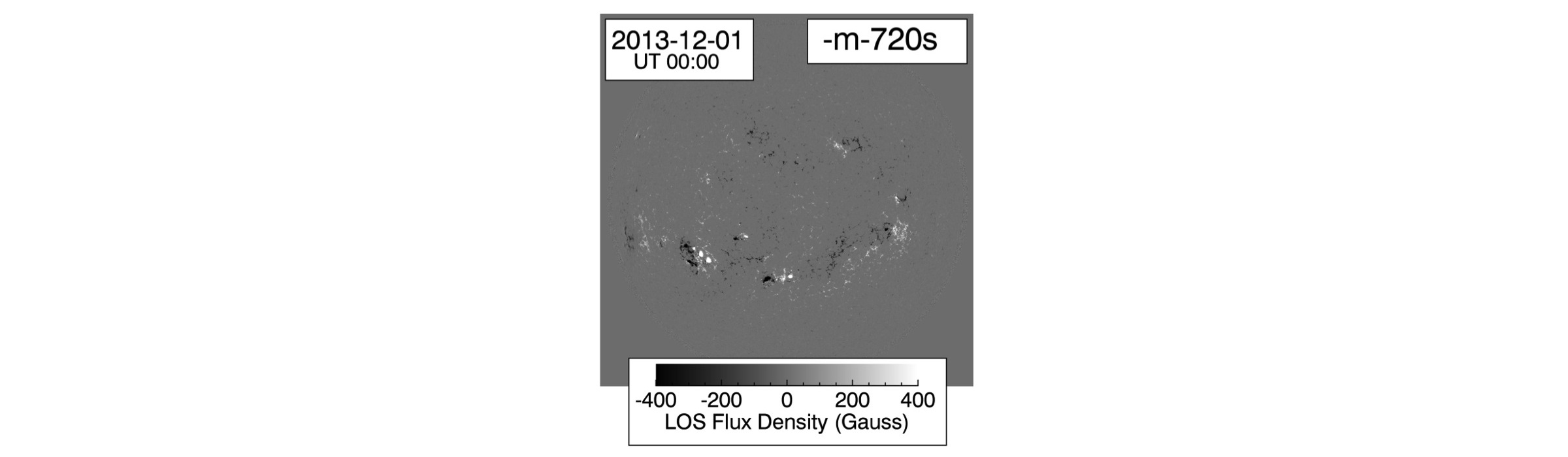}
     \caption[Example LOS Sky Image]{Example image for an HMI-LOS magnetogram taken Dec.~1$^\mathrm{st}$, 2013 at 00:00 UT.}
     \label{FIG_PSk_LOSlev2}
     \end{center}
     \end{figure}

\noindent {\bf LOS magnetograms:} For the LOS magnetograms, an example output image (from step 2) is shown in Figure~\ref{FIG_PSk_LOSlev2}.
These FITS files have only a single image frame, containing the LOS magnetic flux measured by HMI.  They are given file names with the structure:\\

`x4x72{\bf YYMMDD}t{\bf HHMMSS}\_source-SDO-HMI\_type-m-720s.fits.gz' 
\begin{list}{$\circ$}{}
   \item `x4x72': This is the product code that denotes HMI sky images of photospheric longitudinal magnetograms.
   \item `{\bf YYMMDD}t{\bf HHMMSS}': This is the observation's time-stamp.
   \item `\_type-m-720s': This indicates HMI--line-of-sight--magnetogram source data, regardless of image type.
\end{list}

\begin{figure}[h]
     \begin{center}
     \includegraphics[scale=0.19]{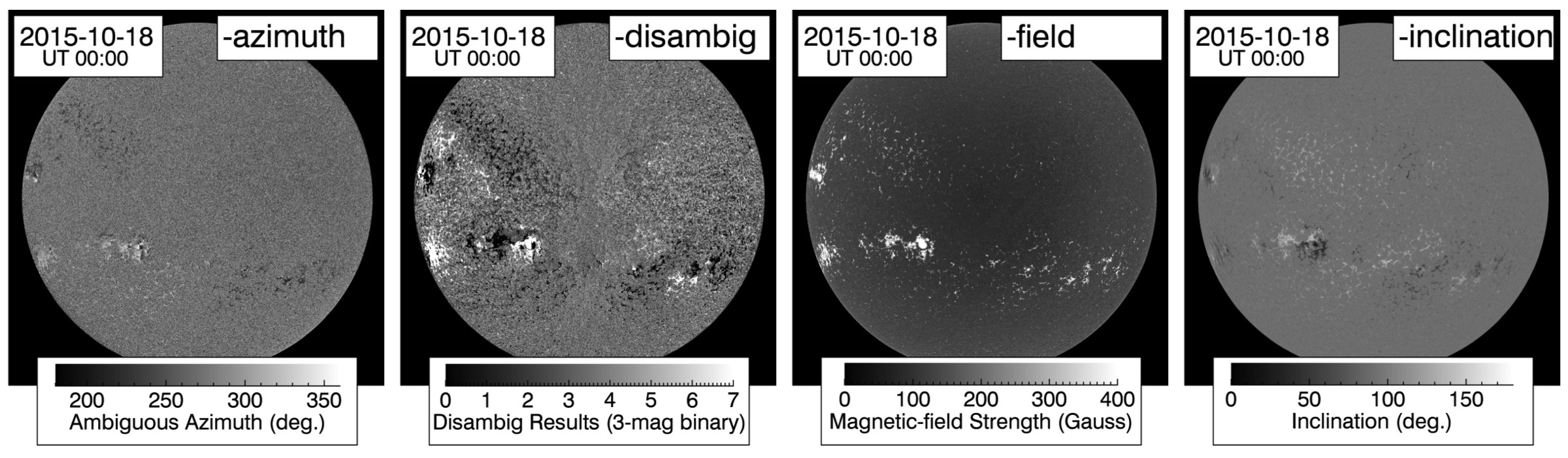}
     \caption[Example Vector-input Sky Image Set]{Set of example input image frames for an HMI-Vector  magnetogram taken Oct.~18$^\mathrm{th}$, 2015 at 00:00 UT.}
     \label{FIG_PSk_Vlev1}
     \end{center}
     \end{figure}

\noindent {\bf Vector magnetograms:} For the vector magnetograms, an example file set of ingested (output from step 2) data are shown in Figure~\ref{FIG_PSk_Vlev1}.  In the `-azimuth' file, angles are measured from the +y image axis and increase counter-clockwise.  In the `-disambig' file, true values for the Radial-acute disambiguation are indicated with integer values 4,5,6 and 7 (for Random disambiguation: 2,3,6,7; for Potential-acute disambiguation: 1,3,5,7).

\begin{figure}
     \begin{center}
     \includegraphics[scale=0.19]{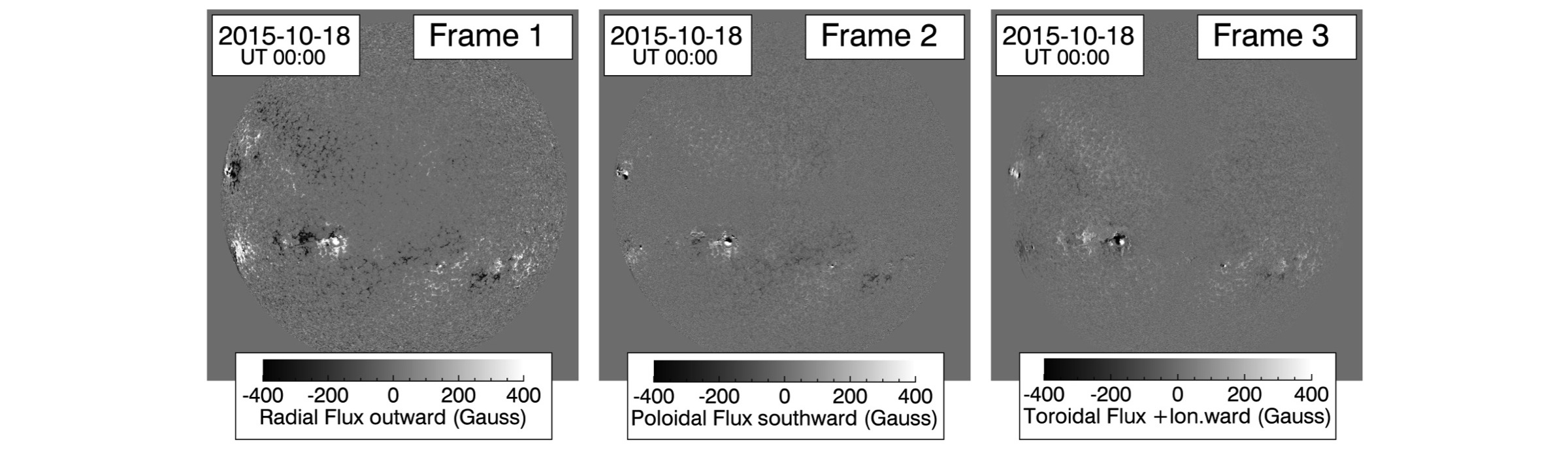}
     \caption[Example Heliocentric-Vector Sky Image Set]{Set of example image frames for an HMI-Vector-derived Heliocentric-vector magnetogram observed Oct.~18$^\mathrm{th}$, 2015 at 00:00 UT.}
     \label{FIG_PSk_Vlev2}
     \end{center}
     \end{figure}

The frames for the corresponding heliocentric--magnetic-vector--components file (output from step 3) are shown in Figure~\ref{FIG_PSk_Vlev2}.  These final-vector sky-image output FITS files have naming structures and frame contents as outlined below.\\



\noindent{\bf Filename Structure:}\\
`x4x92{\bf YYMMDD}t{\bf HHMMSS}\_source-SDO-HMI\_type-b-720s-helio.fits.gz' 
\begin{list}{$\circ$}{}
   \item `x4x92': This is the product code that denotes HMI sky images of photospheric vector magnetograms.
   \item `{\bf YYMMDD}t{\bf HHMMSS}': This is the observation's time-stamp.
   \item `\_type-b-720s-helio': Regardless of image type, this indicates HMI-vector-magnetogram source data mapped into heliocentric vector components.
\end{list}

\noindent {\bf FITS-frame Contents:}\\
$\,$\\
\noindent
\begin{tabular}{cp{0.6in}q{4.9in}}
     {\bf frame} &
        {\bf units} &
        \noindent {\bf title / description}
        \\
     \hline
        &&\\
     1 & Gauss &
        \noindent {\bf Radial flux (outward):} \endgraf
        $\,\,\,\,\,\,\,\,\,$HMI b\_720s magnetogram radial-flux vector component.
        \\
        &&\\
     2 & Gauss &
        \noindent {\bf Poloidal flux (southward):} \endgraf
        $\,\,\,\,\,\,\,\,\,$HMI b\_720s magnetogram poloidal-flux vector component.
        \\
        &&\\
     3 & Gauss &
        \noindent {\bf Toroidal flux (+longitude-ward):} \endgraf
        $\,\,\,\,\,\,\,\,\,$HMI b\_720s magnetogram toroidal-flux vector component.
        \\
        &&\\
\end{tabular}


	\subsection[\textcolor{green}{Heliographic Remaps}]{\textcolor{green}{Heliographic Remaps}}
	\label{Processing_Remaps}

Once the HMI sky images have been prepared for ingest into SOLIS synoptic-map processing (\S\ref{Processing_SkyImages}), the next step is to map each image into a grid of longitude-sine(latitude) heliographic coordinates, as follows:
\begin{enumerate}
\setcounter{enumi}{3}
\item Call {\bf hmi\_serrmaps\_remap\_BatchRun.sh} [-v] {\em START STOP OUTDIR}:
     \begin{list}{-}{}
     \item For each day from {\em START} to {\em STOP}:
          \begin{list}{$\bullet$}{}
          \item Searches for available prepped sky images, {\em SKYFILE}s.
          \item  In IDL, calls {\bf hmi\_serrmaps\_remap}, {\em SKYFILE}, {\em OUTDIR}, /tokeep, /sinlat:
               \begin{list}{{\bf $\,\,\, \rightarrow$}}{}
               \item Reads in the {\em SKYFILE} image frame(s).
               \item Calls {\bf chcoord3.pro} to define the heliographic coordinates of each image pixel, and for all four corners of each pixel.
               \item {\bf IF(longitudinal):}  Projects the line-of-sight flux values into purely radial flux values (as per Equation~\ref{EQ_radproj}).
               \item Defines the Carrington-longitude bounds for the observation to set the bins for the heliographic output map.
               \item Sorts all on-disk image pixels into weighted longitude-sine(latitude) bins.  Pixels that cover multiple heliographic bins may be broken up into as many as 25 (5x5) sub-pixels for heliographic binning.  ({\bf Note:} This matches the spatial resolution of the SOLIS spatial-variance-map sub-pixel binning, where image pixels are broken up into 10x10 sub-pixels but derive from observations of half the spatial resolution as HMI.)
               \item For each heliographic bin, computes:
                    \begin{list}{$\,\,\,\,\circ$}{}
                    \item the sum-of-weights (number of contributing pixels)
                    \item the mean magnetic flux
                    \item the RMS flux variance
                    \item the sum of squared weights
                    \item the mean of squared fluxes
                    \end{list}
                    \begin{list}{$\,\,\,\,\rightarrow$}{}
                    \item {\bf Note:} For {\bf vector magnetograms}, the mean, RMS, and mean-squared fluxes are computed individually for all three vector components.
                    \end{list}
               \item Outputs the resulting heliographic maps of computed quantities into a FITS file using the NSO-style file-naming convention and placed in a data keep in {\em OUTDIR}.
               \end{list}
          \end{list}
     \end{list}
\end{enumerate}

\begin{figure}
     \begin{center}
     \includegraphics[scale=0.19]{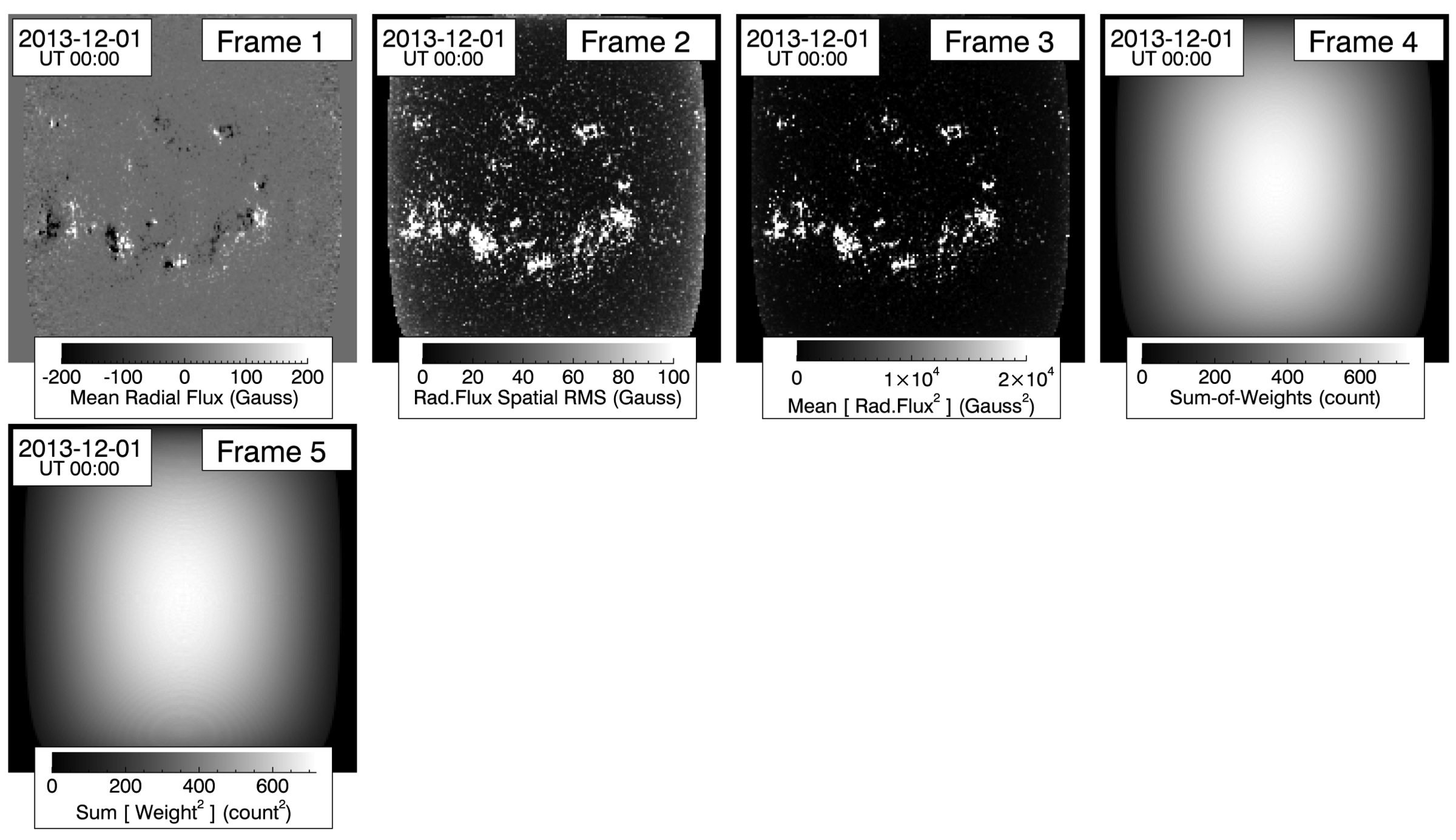}
     \caption[Example LOS Heliographic Remap]{Set of example image frames for an HMI-LOS-derived heliographic remap for an observation taken Dec.~1$^\mathrm{st}$, 2013 at 00:00 UT.}
     \label{FIG_PR_LOSremap}
     \end{center}
     \end{figure}

\noindent {\bf Pseudo-radial Heliographic Maps:}
An example of the frames output for a pseudo-radial heliographic map are shown in Figure~\ref{FIG_PR_LOSremap}.  
These FITS files have naming structures and frame contents as outlined below.\\

\noindent{\bf Filename Structure: }\\
`x9x73{\bf YYMMDD}t{\bf HHMMSS}\_map-err\_dim-180\_source-SDO-HMI\_type-m-720s.fits.gz' 

\begin{list}{$\circ$}{}
   \item `x9x73': This is the product code that denotes HMI heliographic remaps of photospheric longitudinal data.
   \item `{\bf YYMMDD}t{\bf HHMMSS}': This is the observation's time-stamp.
   \item `\_type-m-720s': Regardless of image type, this indicates HMI--line-of-sight--magnetogram source data.
\end{list}


\noindent {\bf FITS-frame Contents:}\\
$\,$\\
\noindent
\begin{tabular}{cp{0.6in}q{4.9in}}
     {\bf frame} &
        {\bf units} &
        \noindent {\bf title / description}
        \\
     \hline
        &&\\
     1 & Gauss &
        \noindent {\bf Weighted-mean Radial Flux Density:} \endgraf
        $\,\,\,\,\,\,\,\,\,$Mean of radially-projected HMI-LOS magnetic flux at each heliographic bin.
        \\
        &&\\
     2 & Gauss &
        \noindent {\bf Spatial RMS Estimate:} \endgraf
        $\,\,\,\,\,\,\,\,\,$Statistical variance of all radially-projected HMI-LOS-flux values at each heliographic bin.
        \\
        &&\\
     3 & Gauss$^2$ &
        \noindent {\bf Mean squared-Radial Flux:} \endgraf
        $\,\,\,\,\,\,\,\,\,$Mean of squared pseudo-radial flux values at each heliographic bin.
        \\
        &&\\
     4 & counts &
        \noindent {\bf Sum-of-Weights:} \endgraf
        $\,\,\,\,\,\,\,\,\,$Sum of weights (image-pixel fractions) into each heliographic bin.
        \\
        &&\\
%
%
%
%
%
     5 & counts$^2$ &
        \noindent {\bf Sum-of-squared-Weights:} \endgraf
        $\,\,\,\,\,\,\,\,\,$Sum of squared-weights into each heliographic bin.
        \\
        &&\\
\end{tabular}

\begin{figure}
     \begin{center}
     \includegraphics[scale=0.19]{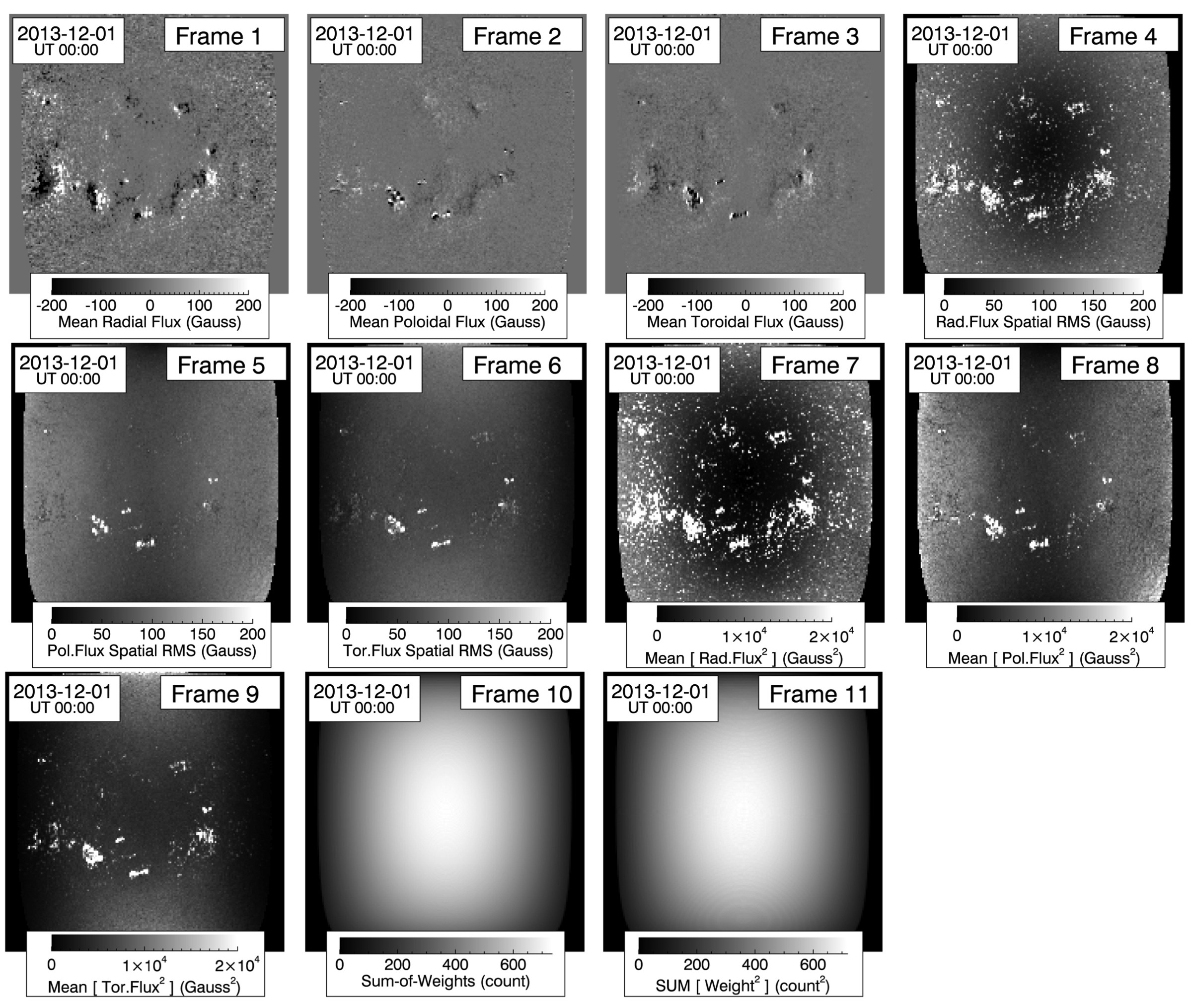}
     \caption[Example Vector Heliographic Remap]{Set of example image frames for an HMI-Vector-derived heliographic remap for an observation taken Dec.~1$^\mathrm{st}$, 2013 at 00:00 UT.}
     \label{FIG_PR_Vremap}
     \end{center}
     \end{figure}

\noindent {\bf Vector Heliographic Maps:}
An example of the frames output for a vector heliographic map are shown in Figure~\ref{FIG_PR_Vremap}.  
These FITS files have naming structures and frame contents as outlined below.\\

\noindent{\bf Filename Structure:} \\
`x9x93{\bf YYMMDD}t{\bf HHMMSS}\_map-err\_dim-180\_source-SDO-HMI\_type-b-720s-helio.fits.gz' 

\begin{list}{$\circ$}{}
   \item `x9x93': This is the product code that denotes HMI heliographic maps of photospheric vector magnetograms.
   \item `{\bf YYMMDD}t{\bf HHMMSS}': This is the observation's time-stamp.
   \item `\_type-b-720s-helio': Regardless of image type, this indicates HMI-vector-magnetogram source data mapped into heliocentric vector components.
\end{list}

\pagebreak

\noindent {\bf FITS-frame Contents:}\\
$\,$\\
\noindent
\begin{tabular}{cp{0.6in}q{4.9in}}
     {\bf frame} &
        {\bf units} &
        \noindent {\bf title / description}
        \\
     \hline
        &&\\
     1 & Gauss &
        \noindent {\bf  Mean Radial (outward) Flux Density:} \endgraf
        $\,\,\,\,\,\,\,\,\,$Mean of HMI-vector radial flux at each heliographic-coordinate bin.
        \\
        &&\\
     2 & Gauss &
        \noindent {\bf Mean Poloidal (southward) Flux Density:} \endgraf
        $\,\,\,\,\,\,\,\,\,$Mean of HMI-vector poloidal flux at each heliographic-coordinate bin.
        \\
        &&\\
     3 & Gauss &
        \noindent {\bf Mean Toroidal (+longitude-ward) Flux Density:} \endgraf
        $\,\,\,\,\,\,\,\,\,$Mean of HMI-vector toroidal flux at each heliographic-coordinate bin.
        \\
        &&\\
     4 & Gauss &
        \noindent {\bf  Radial-flux Spatial RMS Estimate:} \endgraf
        $\,\,\,\,\,\,\,\,\,$Statistical variance of all HMI-vector radial-flux values into each heliographic-coordinate bin.
        \\
        &&\\
     5 & Gauss &
        \noindent {\bf Poloidal-flux Spatial RMS Estimate:} \endgraf
        $\,\,\,\,\,\,\,\,\,$Statistical variance of all HMI-vector poloidal-flux values into each heliographic-coordinate bin.
        \\
        &&\\
     6 & Gauss &
        \noindent {\bf Toroidal-flux Spatial RMS Estimate:} \endgraf
        $\,\,\,\,\,\,\,\,\,$Statistical variance of all HMI-vector toroidal-flux values into each heliographic-coordinate bin.
        \\
        &&\\
     7 & Gauss$^2$ &
        \noindent {\bf Mean squared-Radial Flux:} \endgraf
        $\,\,\,\,\,\,\,\,\,$Mean of squared radial-flux values at  each heliographic-coordinate bin.
        \\
        &&\\
     8 & Gauss$^2$ &
        \noindent {\bf Mean squared-Poloidal Flux:} \endgraf
        $\,\,\,\,\,\,\,\,\,$Mean of squared poloidal-flux values at each heliographic-coordinate bin.
        \\
        &&\\
     9 & Gauss$^2$ &
        \noindent {\bf Mean squared-Toroidal Flux:} \endgraf
        $\,\,\,\,\,\,\,\,\,$Mean of squared toroidal-flux values at each heliographic-coordinate bin.
        \\
        &&\\
     10 & counts &
        \noindent {\bf Sum-of-Weights:} \endgraf
        $\,\,\,\,\,\,\,\,\,$Sum of weights (image-pixel fractions) into each heliographic-coordinate bin.
        \\
        &&\\
     11 & counts$^2$ &
        \noindent {\bf Sum-of-squared-Weights:} \endgraf
        $\,\,\,\,\,\,\,\,\,$Sum of squared-weights into each heliographic-coordinate bin.
        \\
        &&\\
\end{tabular}


	\subsection[\textcolor{green}{Compiling Synoptic Maps}]{\textcolor{green}{Compiling Synoptic Maps}}
	\label{Processing_Synoptic}

Once all of the heliographic remaps have been processed (\S\ref{Processing_Remaps}), they can be assembled into Integral synoptic maps covering the full 360$^\circ$ of Carrington longitude, as follows:
\begin{enumerate}
\setcounter{enumi}{4}
\item Call {\bf hmi\_serrmaps\_synoptic\_BatchRun.sh} [-v] {\em START STOP CARRFILE}:
     \begin{list}{-}{}
     \item Uses {\em CARRFILE} to look up the date ranges of the Carrington rotations, {\em CARRNUM}s.
     \item For each {\em CARRNUM} ocuring between {\em START} and {\em STOP}:
          \begin{list}{$\bullet$}{}
          \item  Calls the IDL routine {\bf hmi\_serrmaps\_synoptic.pro} for the specified {\em CARRNUM} and data type (HMI-LOS or HMI-Vector):
               \begin{list}{{\bf $\,\,\, \rightarrow$}}{}
               \item Looks up the date range covered by {\em CARRNUM} and searches the data keep for a list of all available heliographic remaps falling within that date range +/- an additional 8 days.
               \item Reads in the headers of the listed heliographic files in order to:
                     \begin{list}{$\,\,\,\,${\bf *}}{}
                     \item Define the range of longitude bins covered by each heliographic map.
                     \item Discard from the list any heliographic maps that fall entirely outside the 0-360$^\circ$ longitude of {\em CARRNUM} (e.g., usually discards the maps from observations taken 8 days before and after the Carrington-rotation date bounds).
                     \item Double-check various observation-quality keywords and discard any heliographic maps that fail.
                     \end{list}
               \item For each heliographic-map file, {\em HRFILE}, retained from the file list:
                    \begin{list}{$\,\,\,\,${\bf *}}{}
                    \item Reads in the {\em HRFILE} image frames.
                    \item Rescales the values in the Weights frame by cosine$^4$($\Delta$longitude) with respect to the central meridian.
                    \item Places all in-bounds heliographic-map data into the synoptic-map image space.  In this step, each heliographicly mapped quantity for this observation (weights, fluxes, etc.)~is saved into its own synoptic-map of an {\em nfiles} stacked set.
                    \end{list}
               \item Once all of the heliographic maps have been loaded into the synoptic-map space, computes:
                    \begin{list}{$\,\,\,\,\circ$}{}
                    \item the sum of weights in each synoptic-map bin
                    \item the mean weighted-flux values in each synoptic-map bin
                    \item the spatial variance of the flux values in each synoptic-map bin
                   \end{list}
                    \begin{list}{$\,\,\,\,\rightarrow$}{}
                    \item {\bf Note a:} For {\bf vector maps}, the mean-flux and spatial-variance values are computed individually for all three vector components.
                    \item {\bf Note b:} For any synoptic-map bin where the sum-of-weights equals 0, the mean-flux value(s) is set to 0, and the spatial-variance(s) is flagged with the nonsense value -1000.
                    \end{list}
               \item {\bf IF(HMI-longitudinal):}  Calls {\bf hmi\_serrmaps\_polefiller\_sfit.pro} to return a pole-filled version of the radial-flux map.
               \item Outputs the final synoptic maps of computed quantities into a FITS file using the NSO-style file-naming convention outlined in \S\ref{Product}.
               \end{list}
          \end{list}
     \end{list}
\end{enumerate}

This is the final stage of processing, which produces the data-product files described in \S\ref{Product}.

\section[\textcolor{green}{Notes on HMI Disambiguation}]{\textcolor{green}{Notes on HMI Disambiguation}}
\label{Considerations}

Creation of synoptic maps from HMI magnetograms required a few choices as to the handling of the HMI data, and primary among them was which quiet-sun disambiguation results should be employed to project the observed vector fields into heliographic coordinates.
%

The -disambig file included with all HMI-vector magnetograms in the b\_720s series provides the HMI-disambiguation results as an image of true/false values answering whether the azimuth angle at a given pixel should be rotated by 180$^\circ$ relative to the value provided in the -azimuth file \citep{citeHMI_vectorpipe}.  For strong-field and near-strong-field pixels, the disambiguation is the result of ``annealing" using a minimum-energy algorithm.  For weak-field pixels, the -disambig file provides results from three different disambiguation algorithms:
\begin{enumerate}
   \item A {\bf Potential-acute} algorithm that works to align the field with a potential field extrapolated from the vertical field component.
   \item A {\bf Random} disambiguation assignation.
   \item A {\bf Radial-acute} algorithm that selects the disambiguation that most closely aligns the field in the purely radial direction.
\end{enumerate}

\begin{figure}[h]
     \begin{center}
     \includegraphics[scale=0.20]{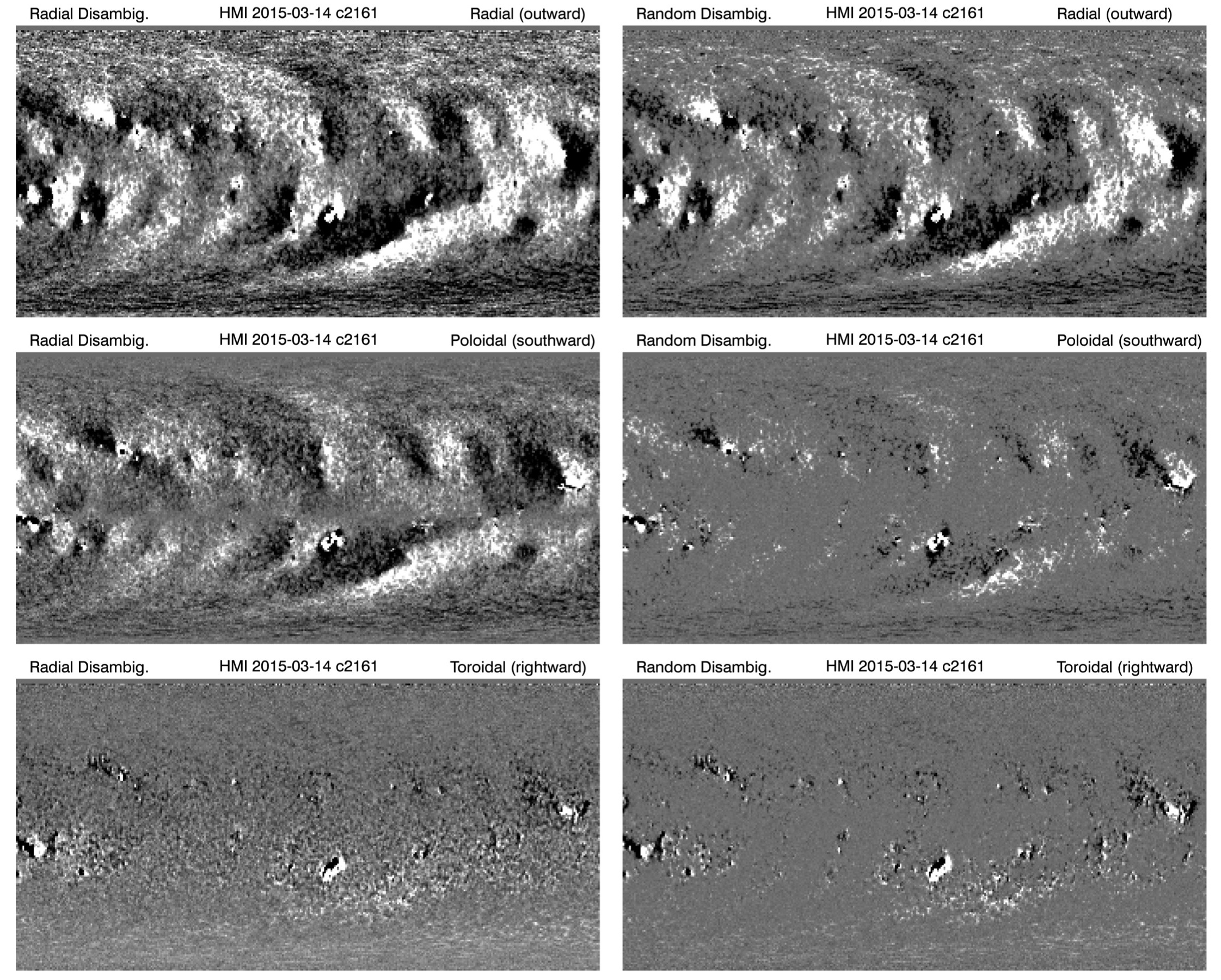}
     \caption[Comparison between Radial-acute and Random Synoptic Maps]{For Carrington rotation 2161 centered around March 14$^\mathrm{th}$ 2015, a comparison of HMI-vector synoptic maps produced using the {\bf Radial-acute} (left column) versus {\bf Random} (right column) weak-field disambiguation results.  From top to bottom, the rows plot the mean-radial, -poloidal, and -toroidal flux, with a greyscale stretching from -20 to +20 Gauss.}
     \label{FIG_Notes_disambigSynoptic}
     \end{center}
     \end{figure}

The HMI documentation \citep{citeHMI_wiki} recommends using \#2, where the weak-field disambiguation is randomly assigned.  However, for these synoptic maps, we have chosen to employ the radial-acute disambiguation results, which produce clearer signatures of the synoptic magnetic field in the quiet sun, as can be seen in Figure~\ref{FIG_Notes_disambigSynoptic}.
     

%
%
%
%


\section*{Acknowledgements}

The authors would like to thank Yang Liu for assistance with the appropriate use of HMI FITS-file header keywords.
%
This work was partially supported by NASA grant NNX15AN43G.

\addcontentsline{toc}{section}{\textcolor{green}{Bibliography}}


\begin{thebibliography}{9}

\bibitem[Bertello, et al.(2014)]{citeBertello2014}
Bertello, L., Pevtsov, A.~A., Petrie, G.~J.~D., and Keys, D.  2014.  ``Uncertainties in Solar Synoptic Magnetic Flux Maps", Solar Physics, {\bf 289}, 2419-2431.

\bibitem[Hoeksema, et al.(2014)]{citeHMI_vectorpipe}
Hoeksema, J.~T., +11 co-authors  2014. ``The Helioseismic and Magnetic Imager (HMI) Vector Magnetic Field Pipeline: Overview and Performance", Solar Physics, {\bf 289}, 3483-3530.

\bibitem[JSOC Wiki - Disambiguation(2014)]{citeHMI_wiki}
``JSOC Wiki - Full-Disk Disambiguated Vector Magnetic Field from HMI." {\em http://jsoc.stanford. edu/jsocwiki/FullDiskDisamb}.  Joint Science Operations Center, May 2014, Web.~8 Apr.~2016.

\bibitem[Leka, et al.(2009)]{citeLeka2009}
Leka, K.~D., Barnes, G., Crouch, A.~D., Metcalf, T.~R., Gary, G.~A., Jing, J., and Liu, Y.  2009. ``Resolving the 180$^\circ$ Ambiguity in Solar Vector Magnetic Field Data: Evaluating the Effects of Noise, Spatial Resolution, and Method Assumptions", Solar Physics, {\bf 260}, 83-108.

\bibitem[Metcalf, et al.(2006)]{citeMetcalf2006}
Metcalf, T.~R., +14 co-authors  2006. ``An overview of existing algorithms for resolving the 180$^\circ$ ambiguity in vector magnetic fields: Quantitative tests with synthetic data", Solar Physics, {\bf 237}, 267-296.

\bibitem[Scherrer, et al.(2012)]{citeHMI_basic}
Scherrer, P.~H., +12 co-authors  2012. ``The Helioseismic and Magnetic Imager (HMI) Investigation for the Solar Dynamics Observatory (SDO)", Solar Physics, {\bf 275}, 207-227.

\bibitem[Schou, et al.(2012)]{citeHMI_instrument}
Schou, J., +20 co-authors  2012. ``Design and Ground Calibration of the Helioseismic and Magnetic Imager (HMI) Instrument on the Solar Dynamics Observatory (SDO)", Solar Physics, {\bf 275}, 229-259.

\bibitem[Sun, et al.(2011)]{citeSun2011}
Sun, X., Liu, Y., Hoeksema, J.~T., Hayashi, K., Zhao, X.  2011. ``A New Method for Polar Field Interpolation", Solar Physics, {\bf 270}, 9-22.

\end{thebibliography}
\end{document}